# Perceptions and worldviews of Transgender individuals

Eiji YAMAMURA[1*][a]

1. *Department of Economics, Seinan Gakuin University, 6-2-92 Sawaraku Nishijin Fukuoka 814-8511, Japan.*

\* Corresponding author

E-mail: yamaei@seinan-gu.ac.jp

a. Conceptualization, Formal analysis, Investigation, Methodology, Writing – original draft preparation.

# Abstract

This study explores the different subjective values held by transgender people, including their subjective well-being, self-reported health status, and career-oriented decision-making. Using an individual-level panel dataset of over 19,000 observations, we discovered the following statistically significant findings: (1) The likelihood of transgender people being happy and healthy is lesser than that of non-transgender people by 7% and 12%, respectively. (2) The likelihood of transgender people supporting women empowerment and giving importance to changing one's behavior for a desirable spouse is 5% lesser than that of non-transgender people. Transgender individuals are also less likely than others to endorse gender-related statements, irrespective of their direction. (3) transgender people are 12% less likely than non-transgender people to make independent decisions for their future career and 2% more likely to follow their parents' and teachers' opinions. (4) Transgender people are 5% more likely to generally distrust others than non-transgender people. Transgender people's subjective well-being and health status outcomes are consistent with those of previous studies, whereas their results for gender-related issues and decision-making do not align with the progressive view.

# 1 Introduction

Despite having its benefits, disclosure of one's transgender identity may exacerbate harm, which is why transgender people often consider the possible costs and benefits of disclosing their identities (Beagan et al., 2023). For those who decide to come out, the benefits of disclosure may outweigh the costs. Media coverage of transgender individuals has steadily increased over the last 10 years (Ng et al., 2024). Moreover, the existence of transgender people is increasingly being acknowledged in society. However, compared with non-transgender people, transgender individuals are more likely to experience hate crimes (Herek, 2009; Timmins et al., 2017), problems in their social lives, and negative emotional responses (Flores et al., 2022). Owing to such environmental circumstances, transgender individuals' health status and subjective well-being (SWB) have deteriorated (Descamps et al., 2000; Duncan & Hatzenbuehler, 2014; Editorial, 2024; Herek et al., 1999; McCabe et al., 2010; Russell & Fish, 2016). Furthermore, the probability of transgender people committing suicide has increased (Clements-Nolle et al., 2006; Duncan & Hatzenbuehler, 2014).

Nearly 70% of transgender people avoid disclosing their gender identity for fear of experiencing negative reactions from others (Government Equalities Office, 2018). Those who have come out are more likely to be involved in social activism and tackling gender issues. Transgender people's subjective views about society may inevitably suffer from selection biases. Compared with transgender people who have already disclosed their identity, closeted transgender people are thought to have different views about concealing theirs.

Previous studies have provided evidence for transgender people's social characteristics (Carpenter et al., 2020; Government Equalities Office, 2018; Spizzirri et al., 2022), views, and health status (Editorial, 2024; Government Equalities Office, 2018; Russell & Fish, 2016; Scandurra et al., 2021; Scheim et al., 2024; Wahlen et al., 2020). Transgender individuals tend to experience higher rates of self-reported psychological and verbal violence than others (Flores et al., 2022; Herek, 2009; Rothman et al., 2011; Spizzirri et al., 2022). Violent crimes may occur when they reveal their transgender identity. In contrast, the sexuality of closeted transgender people is difficult to determine based on their appearance and attitude, particularly if they succeed in pretending to behave like non-transgender people. Because of this, previous studies have not been able to examine the SWB, health status, views, and perceptions of closeted transgender people.

In this study, without asking about the respondents' transgender identity, we identified transgender people as those who had changed their gender in the questionnaire during the survey period from 2016 to 2024. This novel way of identifying transgender people enabled us to attenuate selection biases when examining the subjective views of transgender people.

Using this identification strategy, we observed that approximately 0.7% of the adult population comprised transgender people, a finding which is consistent with that of previous studies that indicated a rate of 0.1–2.0% (Goodman et al., 2019; Spizzirri et al., 2021). To explore transgender people's views, we used individual-level panel data for the period 2016–2024 to identify those who had changed their gender as transgender.

To truly understand different transgender individuals and their selection biases, we must consider their perceptions and views by asking several questions. However, studies have generally included small sample sizes in examining transgender people's subjective views (Henriquez & Ahmad, 2021), and the statistical significance of their results has not been sufficiently verified (Government Equalities Office, 2018). The purpose of this study was to examine how transgender individuals' subjective views differ from non-transgender individuals' views by attenuating selection biases and using a large sample. Our major findings reported that transgender people have a more negative view of gender-related statements than non-transgender people, regardless of the conservative nature of the statement. Additionally, transgender people may follow their teachers' and parents' opinions when deciding their future career direction. However, they are less likely to trust others.

This study demonstrates that the views of transgender people do not necessarily coincide with the stereotypes described by social activists. The impression that transgender individuals are radically progressive may be attributed to selection bias. Those who do not publicly identify as transgender seem to have a mindset that cannot be divided into the existing dichotomous categories of conservative and progressive.

## 2 Data

### 2.1 Data collection

Since 2016, we have been compiling panel data using a long-term internet survey called "The Survey on Childhood Living Environments and Current Life Attitudes." Questions related to the COVID-19 pandemic were included in the 2020 survey, resulting in a specific title for this project: "A Study on the Influence of the New Type of Coronavirus Infection on Lifestyle Consciousness." The data for this survey were collected in February 2024, when the COVID-19 pandemic had seemingly concluded.

We conducted internet surveys to gather individual-level data. The Nikkei Research Company (NRC), a research firm having abundant experience with academic surveys, was selected to conduct the project. We attempted to pursue the same respondents throughout 2016, 2017, 2018, 2021, 2023, and 2024. Respondent withdrawal is a common problem in panel surveys. To maintain an adequate sample size for statistical analysis, new participants were added to the sample.

Most questions were consistently included throughout the study. However, several questions were replaced with new ones to explore the respondents' various subjective values. The NRC sent questionnaires to the participants and collected the data. They continued to administer questionnaires until a sufficient sample size was reached. We obtained over 5,000 observations from the survey annually. The final sample consisted of 47,195 respondents. However, because we could not obtain some participants' responses to several questions related to the variables used in this study, the maximum number of analyzed observations was 32,330. Nevertheless, the sample size varied according to specifications.

The questionnaires included questions regarding subjective values, such as SWB and self-reported health status, and demographic information, such as job status, gender, birth year, household income, and residential prefecture. This information was collected from all the surveys.

In previous studies on transgender concerns, transgender individuals were identified by self-statements in a questionnaire that purposefully examined their characteristics and subjective values. However, only self-affirming transgender individuals could disclose their attributes. Consequently, they might have had strong intentions to change society and realize an ideal society for themselves. However, the estimation results of previous studies may have suffered from selection biases.

To mitigate the selection biases, we used the novel strategy that has not been employed in previous works dealing with issue of gender minorities. Liebler et al. (2017) proposed the innovative strategy to consider identity change. Around 6 % in the U.S.

Census Bureau data have a different race and/or Hispanic-origin response in 2010 in compared with they did in 2000 (Liebler et al. 2017). This indicates that information of respondents changing their own identity is useful in the field of social science. Applying the way of Liebler et al. (2017) to transgender issues, we identified some participants who had changed their gender between different survey years, although they constituted approximately 0.07% of the sample and were therefore extremely rare. Initially, we considered these cases as entry errors that may have occurred when completing the online survey because these respondents had inadvertently made several typing errors in other responses. However, participants who had changed their response for gender did not change it for their birth year. Despite the question not being explicitly related to transgender concerns, this response pattern indicated that they intentionally changed their gender. Accordingly, we identified transgender individuals as those who had changed their gender in the questionnaire during the survey period (2016–2024). It is believed that these individuals naturally express their gender identity without experiencing mental stress.

## 2.2 Basic statistics

Table 1 presents some basic information on the key and control variables used in the estimation, such as the definitions of their mean, minimum, and maximum values, and standard errors. Fig 1 illustrates the use of the full-sample data with 32,330 observations. The observations for each dependent variable were used for regression estimations in Figs 2–7. The observations varied between 19,232 and 32,330 according to the key dependent variables for two reasons. (1) The questionnaires in 2021, 2023, and 2024 did not include some of the questions (e.g., those on improving one's manners to be with a desirable spouse, following parents' opinions about career, and making independent decisions about career), and the 2018 questionnaire did not include questions on women's involvement. (2) Some respondents did not answer all questions even when they were included in the questionnaire.

**Table 1. Definitions of Key Variables and Their Basic Statistics.**

| Variables | Definition | Mean | S.D. | Min. | Max. | Obs. |
|---|---|---|---|---|---|---|
| TRANSGENDER | The value is 1 if respondent changed their gender and 0 otherwise. | 0.007 | 0.07 | 1 | 5 | 32,330 |

| **Satisfaction** | How satisfied are you with your life? Please choose one from the five options (1, 2, 3, 4, 5); a higher value indicates a favorable view. | 3.36 | 1.18 | 1 | 5 | 29,706 |
|---|---|---|---|---|---|---|
| **Health** | How is your health condition? Please choose one from the five options (1, 2, 3, 4, 5); a higher value indicates a favorable view. | 3.51 | 1.17 | 1 | 5 | 29,758 |
| **Women empowerment** | "The government should create a society in which women can fully demonstrate their abilities and play an active role in their professional lives." Please choose one from the five options (1, 2, 3, 4, 5); a higher value indicates a favorable view. | 3.38 | 1.18 | 1 | 5 | 24,772 |
| **Improving manners to be with a desirable spouse** | "It is important to improve one's manners and behavior in order to be with a desirable spouse." Please choose one from the five options (1, 2, 3, 4, 5); a higher value indicates a favorable view. | 3.51 | 0.87 | 1 | 5 | 19,232 |
| **Following parents** | "I have followed the opinions of my parents and teachers in making decisions about higher education, employment, etc." Please choose one from the five options (1, 2, 3, 4, 5); a higher value indicates a favorable view. | 2.55 | 1.09 | 1 | 5 | 19,283 |
| **Independent decision** | "Decisions regarding higher education, employment, etc. have been made independently and on one's own initiative." Please choose one from the five | 3.96 | 0.99 | 1 | 5 | 19,283 |
| | options (1, 2, 3, 4, 5); a higher value indicates a favorable view. | | | | | |
| **General trust** | "People are generally trustworthy." Please choose one from the five options (1, 2, 3, 4, 5); a higher value indicates a favorable view. | 3.20 | 1.07 | 1 | 5 | 29,804 |
| **Income** | Household income (million yen):For categories 2–11, mid-values are incorporated. For instance, 30 for >, = 20 and < 40. The lowest category is assumed to be 5 and the highest 23. Respondent's choices are as follows: <br> 1. < 1 million yen <br> 2. > = 1 million yen and < 2 million yen <br> 3. > = 2 million yen and < 4 million yen <br> 4. > = 4 million yen and < 6 million yen <br> 5. > = 6 million yen and < 8 million yen <br> 6. > = 8 million yen and < 10 million yen <br> 7. > = 10 million yen and < 12 million yen <br> 8. > = 12 million yen and < 14 million yen <br> 9. > = 14 million yen and < 16 million yen <br> 10. > = 16 million yen and < 18 million yen <br> 11. > = 18 million yen and < 20 million yen | 6.50 | 4.41 | 0.5 | 23.0 | 32,330 |

|  | 12. >= 20 million yen<br>In regression estimation, household income dummies are included as independent variables.<br>Reported values for household income are calculated based on the household income as follows: The midpoint of the interval was used as the approximate value of household income; that is, the interval of 2–4 million yen was converted to 3 million yen. However, we could not calculate household income for the highest income group of more than 20 million yen. Therefore, we arbitrarily set this level at 23 million yen. |  |  |  |  |  |
|---|---|---|---|---|---|---|
| **Age** | Respondent's age | 47.3 | 12.1 | 18 | 73 | 32,330 |
| **Marriage** | The value is 1 if respondent is married and 0 otherwise. | 0.07 | 0.09 | 0 | 1 | 32,330 |
| **Remarriage** | The value is 1 if respondent is re-married and 0 otherwise. |  |  | 0 | 1 | 32,330 |
| **University** | The value is 1 if respondent has graduated from university and 0 otherwise. | 0.55 | 0.49 | 0 | 1 | 32,330 |

**Fig 1. Differences in Characteristics of TRANSGENDER Individuals.**

**Fig 2. TRANSGENDER Individuals' Subjective Well-Being.**

**Fig 3. TRANSGENDER Individuals' Self-Reported Health Status.**

**Fig 4. TRANSGENDER Individuals' Views on Changing Manners to Find a Desirable Spouse.** Note: Response to the statement "It is important to improve one's

manners and behavior in order to be with a desirable spouse."

**Fig 5. TRANSGENDER Individuals' Views on Women's Involvement.** Note: Response to the statement "The government should create a society in which women can fully demonstrate their abilities and play an active role in the working environment."

**Fig 6A. TRANSGENDER Individuals' Career Decision Being Independent.** Note: Response to the statement "Decisions regarding higher education, employment, etc. have been made independently and on one's own initiative

**Fig 6B. TRANSGENDER Individuals' Career Decision Depending on Teachers' and Parents' Opinions.** Note: Response to the statement "I have followed the opinions of my parents and teachers in making decisions about higher education, employment, etc."

**Fig 7. TRANSGENDER Individuals' General Trust.** Note: Response to the statement "People are generally trustworthy."

As shown in Table 1, the following dummies were included in the set of independent variables in all estimations: 18 types of occupation dummies, 5 dummies for survey years, and 46 residential prefecture dummies. Household income data was collected by asking respondents to choose from 13 interval options. Based on this information, 12 dummy variables for income intervals were constructed and included in the set of independent variables. The reported mean values of household income were calculated using the midpoint in each income interval.

## 2.3  Ethical considerations

Before starting the survey, we obtained consent from all participants. On logging into the online survey site, participants could see a description of the purpose and specifics of the study on the screen. All participants were requested to click the "Start" button to consent to participate in the survey. No individual names were included in the consent form because of the legal requirements to protect personal information. Even after having started with the survey, participants could quit at any time. The authors did not obtain any personal participant information from the NRC. For those participants who were over the age of 18, it was not necessary to obtain consent from their parents or guardians as they were considered adults by Japanese law.

The survey design was approved by the Ethics Committee of the Graduate School of Economics, Osaka University (approval no.: R51127) in November 2023. We began conducting the survey in February 2024. Similarly, we obtained approval from previous surveys to conduct the survey. Data collection was performed in accordance with the relevant guidelines and regulations. This study did not involve experimental manipulation or any foreseeable risks, and the questionnaire used did not elicit any personally identifying information.

The data used in this study are part of long-term panel data on lifestyle attitudes that include a survey of new coronavirus infections. The application documents for ethics review clearly state the purpose of the study as follows: "To accumulate basic knowledge for more desirable policy formulation based on changes in society." Data related to the present study were also collected with this purpose in mind. Therefore, the goal of this study was consistent with the project title "The Survey on Childhood Living Environments and Current Life Attitudes," which was indicated as a subtitle of the survey for ethics approval. However, the main tile was "A Study on the Influence of the New Type of Coronavirus Infection on Lifestyle Consciousness."

# 3 Method

The transgender identity has become stigmatized owing to its negative social evaluation, leading to poor mental health outcomes (Liu et al., 2023; Scandurra et al., 2018; Timmins et al., 2017). Because they are in the minority, transgender people experience a stressful life and are at a high risk for negative health outcomes (Scandurra

et al., 2021).

As indicated in Table 1, the mean value of transgender people in the sample was approximately 0.007, indicating a 0.7% proportion of transgender people in the dataset. This aligns with the findings of previous studies that suggested a rate of 0.1–2.0% (Goodman et al., 2019; Spizzirri et al., 2021).

To quantitatively explore how transgender individuals' views differ from those of the rest of the population, we used ordered logit regression analysis because all dependent variables comprised discrete ordered values between 1 and 5, as indicated in Table 1. We used Stata 14 to automatically indicate the coefficient of the independent variables. However, unlike in the linear model, the coefficient's degree could not be appropriately interpreted because it is not a marginal effect. Hence, we calculated the marginal effect for each dependent variable. We obtained five different marginal effects based on the probability of respondents choosing 1, 2, 3, 4, or 5 for questions regarding the dependent variables.

For SWB, the marginal effect for the probability of choosing 5 indicates that the respondent was very satisfied. However, the marginal effect of choosing 1 indicates the opposite in that the respondent was dissatisfied with their life.

The estimated function takes the following form:

$$Y_i = \alpha_0 + \alpha_1 LGBT_i + X_i A_i + e_i$$

$Y_i$ is the dependent variable, and α denotes the marginal effect of variables. α has 5 different values in an estimation because of the probability that $Y_i$ has 1, 2, 3, 4, and 5. $i$ represents individuals, $X$ is the vector of the control variables, and $A$ is the vector of their marginal effects. $X$ represents marital status, household income, age, educational background, year, and residential prefecture dummies. $e_i$ is an error term. Figs 2–7 demonstrate only the marginal effects for TRANSGENDER individuals because of our focus on their views and perceptions.

# 4 Results

## 4.1 Characteristics of TRANSGENDER individuals

Before the regression estimation results, we examined the characteristics of transgender people. Fig 1 shows a comparison of the percentage of transgender individuals' responses on several dummy variables. For example, those who had changed their residential prefecture during the study period answered "Yes" for the migrant field and "No" if not. In Fig 1, the proportion of transgender individuals being migrants is 1.5% higher than the proportion of non-migrants. We interpret that transgender people may be more likely to feel stressed in their interpersonal relationships in residential places and thus feel motivated to migrate.

Concerning marital status, there was no difference in the percentage of married and unmarried transgender participants. However, the proportion of transgender people who remarried was approximately 2% higher than the proportion of those who did not. This may be because they remarried to fit into their transgender identity after learning from the collapse of their traditional marriage.

Compared with the proportion of transgender people who did not graduate from university, the proportion of those who did is slightly but significantly higher. We suppose that more intelligent people might be able to disclose their transgender identity to the people surrounding them, who may be less likely to adhere to traditional views.

The results shown in Fig 1 indirectly indicate that our study's transgender identification strategy was reliable.

## 4.2 Subjective well-being and self-reported health status

Fig 2 illustrates the marginal effects of transgender status on SWB. We observed that transgender people are more likely to choose "unhappy" or "slightly unhappy" and less likely to choose "happy" or "slightly happy." Their absolute values for the options "happy" and "unhappy" were 0.07 and 0.04, respectively. This means that transgender people are 7% less likely to be "happy" and 4% more likely to be "unhappy" compared with non-transgender people. These values are significantly different from 0. Overall, transgender individuals' SWB is lower than that of the rest of the population.

In Fig 3, transgender people display a higher probability of choosing "unhealthy" or "slightly unhealthy" and a lower probability of choosing "healthy" or "slightly healthy." Their absolute value for unhealthiness was 0.11, which is a distinctly large value. This implies that transgender people are 11% more likely to report their health status as unhealthy than others, a finding that is consistent with previous findings [10,11,13,16–18].

## 4.3  Views on gender-related issues

Fig 4 demonstrates the marginal effects of TRANSGENDER individuals on the importance of the manner of being with a desirable spouse. The probabilities of choosing "disagree" and "slightly disagree" showed significantly different positive values, and those for choosing "agree" and "slightly agree" showed significantly different negative values. Hence, transgender people are more likely to have a negative view of traditional norms. Their absolute values for "agree" and "slightly agree" were approximately 0.5. Thus, transgender people may be 7% less likely than others to hold favorable views of traditional norms.

In Fig 5, we observed the marginal effects of transgender status on women's involvement. Surprisingly, transgender people are more likely to choose "disagree" or "slightly disagree" and less likely to choose "agree" or "slightly agree." These results were significantly different. Their absolute values were 0.02 and 0.03, which implies that, compared with non-transgender people, transgender people are 2–3% less likely to view women's involvement positively. Contrary to the existing stereotypes, transgender people seem to be quite conservative. This may be because transgender people resist questions on traditional dualism, such as that of male and female, because they cannot simply be classified into one of two categories. Transgender individuals' viewpoints may differ from the traditional progressive views of gender issues.

In Fig 6A, concerning individual decision-making about their future career path, we observed that respondents disagreed with the statement "Decisions regarding higher education, employment, etc. have been made independently and on one's own initiative."

However, the probability of agreeing was approximately 0.105, indicating that transgender people are 10.5% less likely to "disagree" than others. The degree of this difference is sizable.

For the statement "I have followed the opinions of my parents and teachers in making decisions about higher education, employment, etc.," transgender people show a lower probability of choosing "disagree" and "slightly disagree" and a higher probability of choosing "agree" and "slightly agree." These values are significantly different between the two groups. Transgender people are 8% less likely to disagree compared with others. Overall, as children, transgender people tend to follow the opinions of the adults surrounding them even for their future. This implies that closeted transgender people hide their true nature and intentions.

## 4.4 General trust

For general trust, Fig 7 demonstrates negative values for the options of "trust" and "slightly trust" and positive values for those of "distrust" and "slightly distrust." These values are significantly different between the two groups. The absolute values for transgender people are approximately 0.05, which implies a 5% higher likelihood for choosing "distrust" and a 5% lower likelihood for choosing "trust." Figs 6 and 7 jointly reveal that transgender people follow parents and teachers but do not trust them. They conceal their true intentions and nature to follow people they do not trust.

# 5 Discussion

## 5.1 Implications

Transgender people are in the minority in society and are therefore thought to be more progressive than the rest of the population. Surprisingly, our findings show that Transgender people disagreed with women's involvement in society. However, they rejected traditional values emphasizing the importance of women having to learn social manners to get married. These results appear to contradict each other.

From a different perspective based on gender dualism, transgender people may be considered the opposite of feminists. Transgender people's thought processes may not fit the old, progressive thinking framework. Lectures and self-assessment tools pertaining to attitudes toward transgender people may be effective in increasing their knowledge and needs in certain situations, such as healthcare services (Kanakubo et al., 2024; Wahlen et al., 2020). Caution should be exercised in representing each subgroup of the transgender community adequately and carefully rather than describing the TRANSGENDER identity as monolithic, as pointed out by some studies (Ng et al., 2024).

Transgender individuals who express their gender identity are more likely to be involved in left-wing social movements. Therefore, the transgender argument tends to be simplified and biased within the old, progressive framework.

Furthermore, closeted transgender people may have more diverse and complicated views than previously believed.

## 5.2 Contributions of the study

Mental health status and SWB of sexual minorities are found to be lower than majorities (Descamps et al., 2000; Duncan & Hatzenbuehler, 2014; Editorial, 2024; Herek et al., 1999; McCabe et al., 2010; Russell & Fish, 2016). Consistent with it, more recently work of Bartman (2023) also provided evidence that sexual minorities such as Gay and Lesbian are less satisfied with their lives than heterosexual persons. The study provided supporting evidence of these studies. Focusing on subjective well-being is too limited to examine how sexual minorities' subjective values are different from majorities. Therefore, extending the object of analysis to various subjective values, this study explore how transgender people's value in their life. At least, as opposed to stereotype, transgender people are found to be self-contradicted in the real situation.

We identified transgender people as those who had changed their gender in the questionnaire during the survey period 2016–2024. We considered this an implicit expression of their gender. In the survey, we frequently provided opportunities for the respondent to answer questions about their gender. They had the right to express their gender without it being related to political matters. When supported by social activists, transgender people may not necessarily have a progressive mindset. Therefore, some transgender people might avoid expressing their sexuality in ideology-oriented questions. Even if they disclosed their identity, they might hide their true intentions in order to meet the progressive group's partisan expectations.

The transgender individuals identified in this study were likely to be hesitant and confused about declaring their true gender, making it difficult to determine how closeted transgender individuals feel and think. In this study, the results of transgender individuals showing lower SWB and self-reported health status were consistent with those of previous studies. However, other findings implied that transgender people follow the directions of their parents and teachers and do not make decisions on their own, despite generally not trusting others. Closeted transgender people are motivated to follow the traditional norms of the majority to avoid discord and conflict with others, even if they are dissatisfied with them. The behavior of closeted transgender individuals is determined by considering its costs and benefits.

In the survey, respondents were not forced to choose a socially desirable answer. The contribution of this study was to elucidate the views and perceptions of transgender individuals who may be uncertain about their gender identity.

## 5.3 Limitations

Despite the large sample size of this study, observations from the transgender sample are few because they occupy only 0.7% of the sample. Besides, the change in respondents' reported gender may be because of typographical errors. If so, the error should occur randomly, and transgender people's characteristics and views should not differ from those of others. However, we observed that transgender people exhibited remarkably different characteristics and subjective views from others. Further, the estimation results of SWB in this study are consistent with previous studies (e.g., Descamps et al., 2000; Duncan & Hatzenbuehler, 2014; Editorial, 2024; Herek et al., 1999; McCabe et al., 2010; Russell &

Fish, 2016; Bartram 2023). It cannot be considered a coincidence that the validation results were similar to those of previous studies.

In addition, changed of own identity is useful information in empirical social studies (Liebler et al. 2017). This demonstrates the validity of our transgender identification strategy. Strictly saying, existing works considering LGBT suffered inevitably selection biases because they are minority. Despite being weakness of identifying transgende, this study did not suffer the biases. Actually, any measure of transgender (LGBT) did not avoid selection and measurement biases. Therefore, comparing various type of studies is important.

Meanwhile, several closeted transgender people may have been present in the sample because of attempts to conceal their identity in the survey. Family acceptance of transgender people has improved the health status of young transgender individuals (Ryan et al., 2010). However, some transgender people may not have disclosed their gender identity to meet the expectations of their parents and others. Therefore, selection biases seem to persist, even in this study.

The boundaries between transgender and non-transgender people may not be clearly drawn. Moreover, the transgender identity may itself be a relative entity. We hope to address these issues in future studies.

# 6 Conclusion

Our estimation results were partially inconsistent. Closeted transgender people might waver between identities. A wider variation may exist in the subjective views and perceptions of different transgender individuals compared with those of non-transgender individuals. To depict the overall picture of transgender issues, we must explore the views and perceptions of closeted transgender people who unconsciously follow traditional norms to conceal their identity.

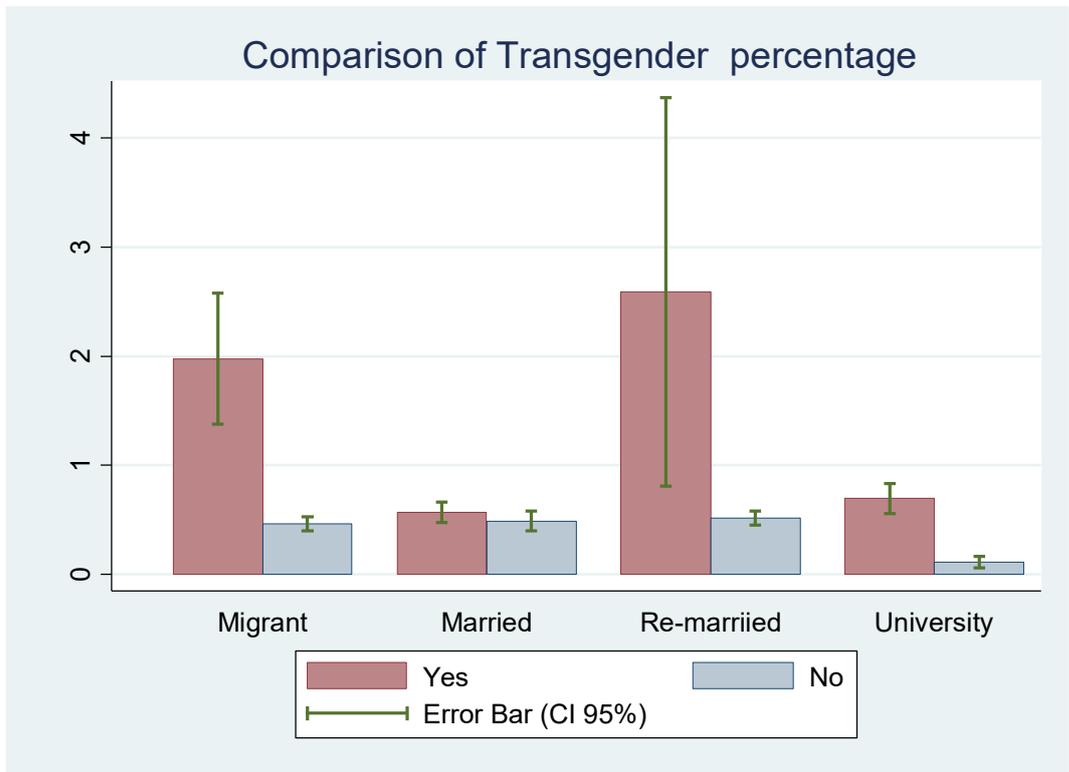

Fig. 1. Difference of characteristics between TRANSGENDER and Non-TRANSGENDER

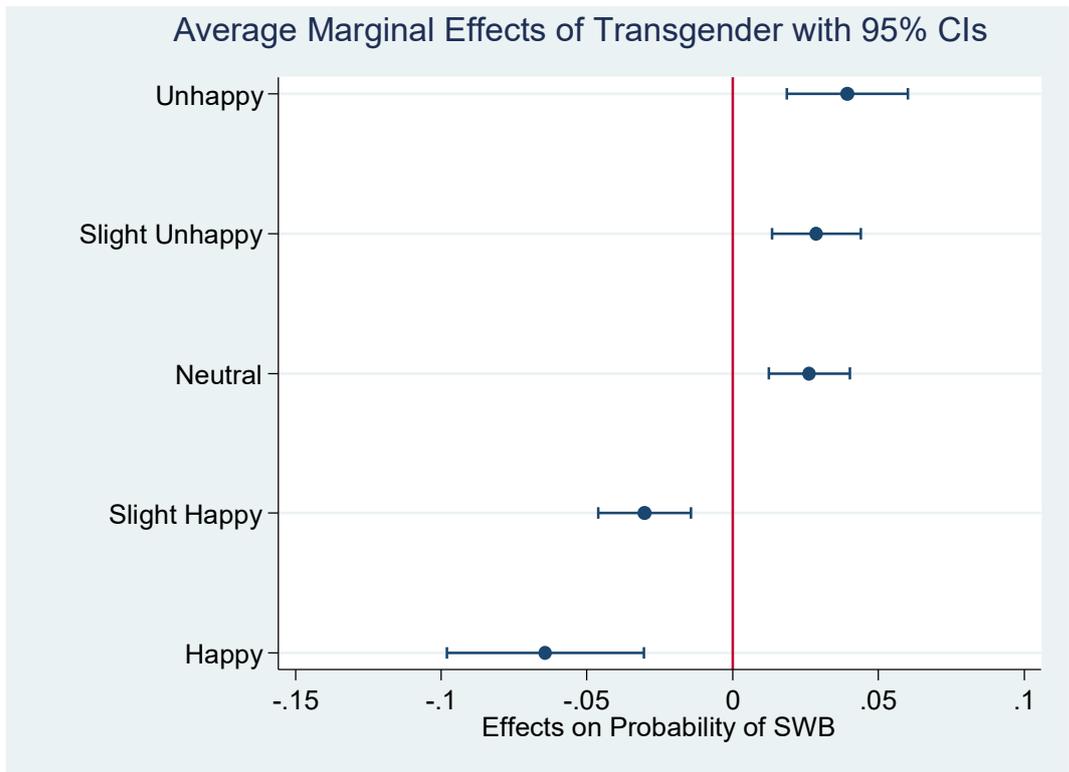

Fig.2.　Influence of TRANSGENDER on SWB

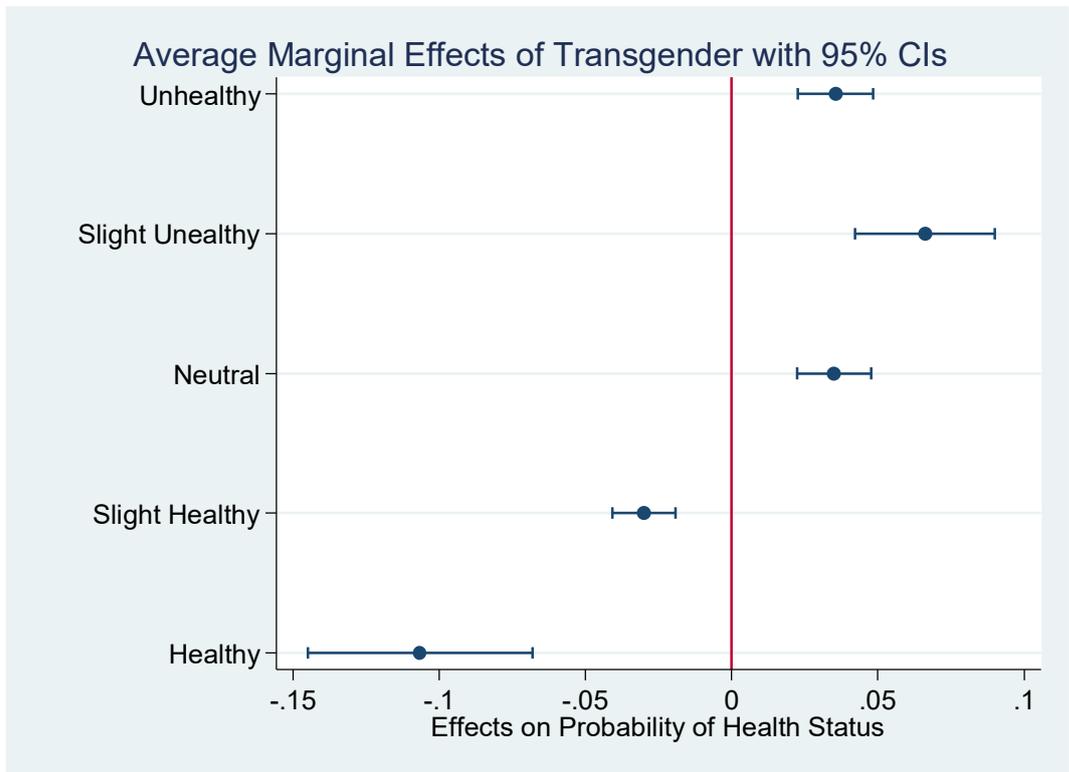

Fig.3    Influence of TRANSGENDER on Stated Health Status

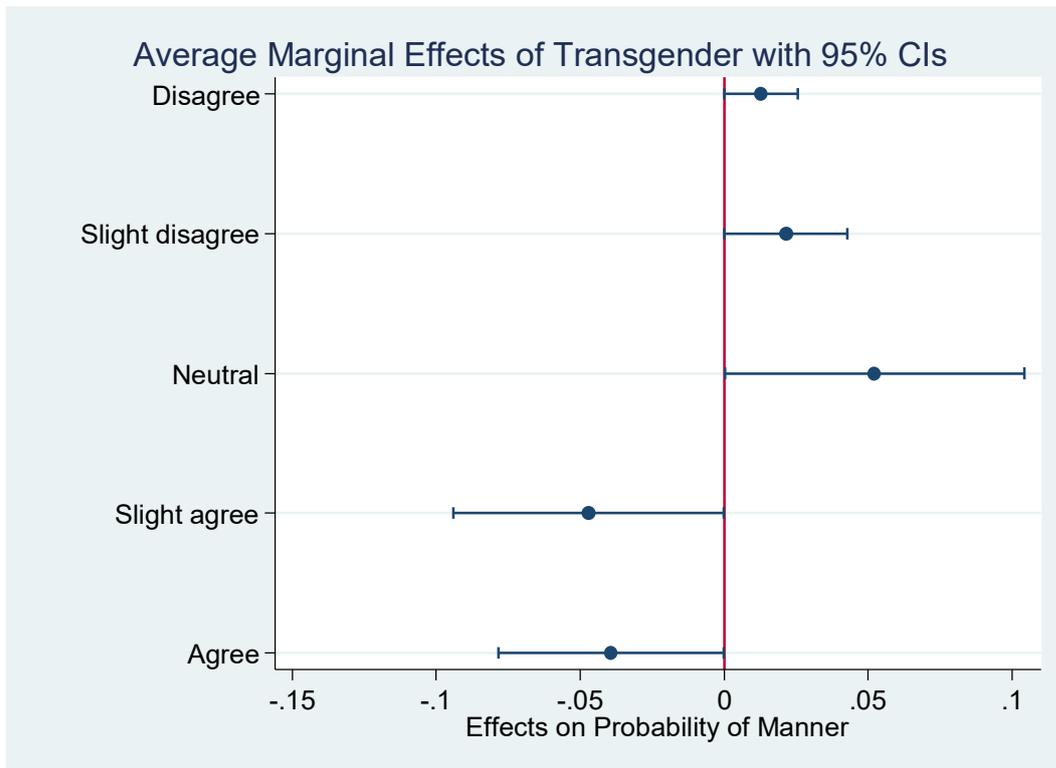

Fig4. Influence of TRANSGENDER on view about importance to get married

Note: Response to the view "It is important to improve one's manners and behavior in order to be with a desirable spouse."

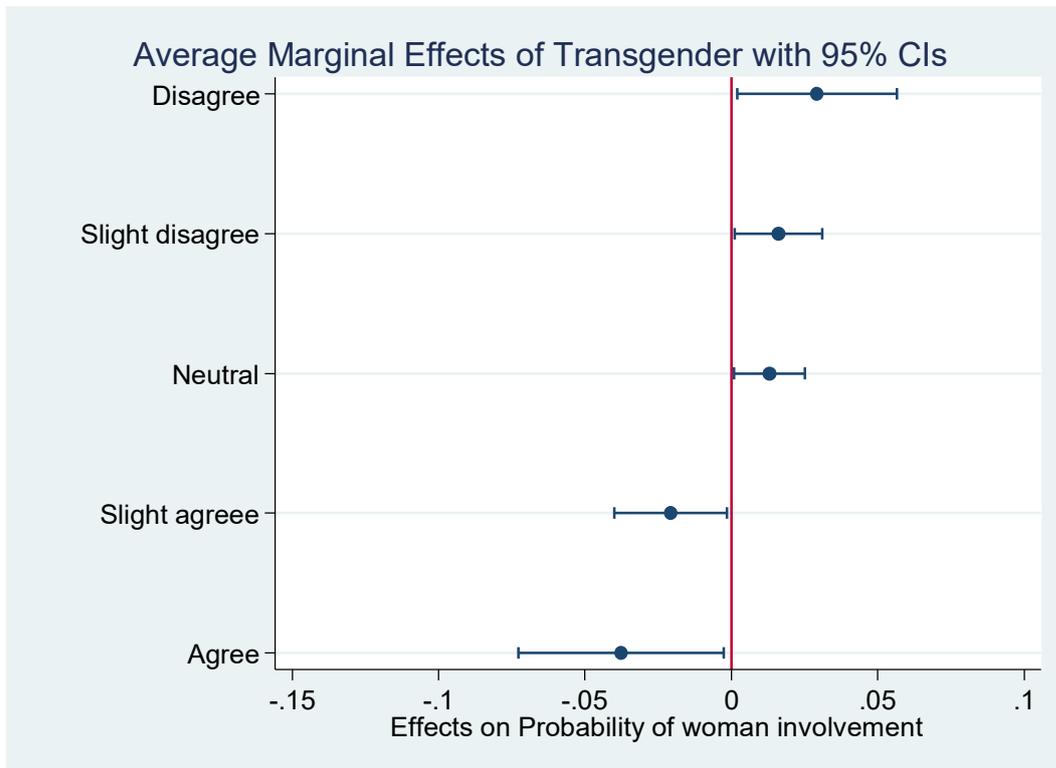

Fig5. Influence of TRANSGENDER on view about women involvement.

Response to the view "The government should create a society in which women can fully demonstrate their abilities and play an active role in the working environment."

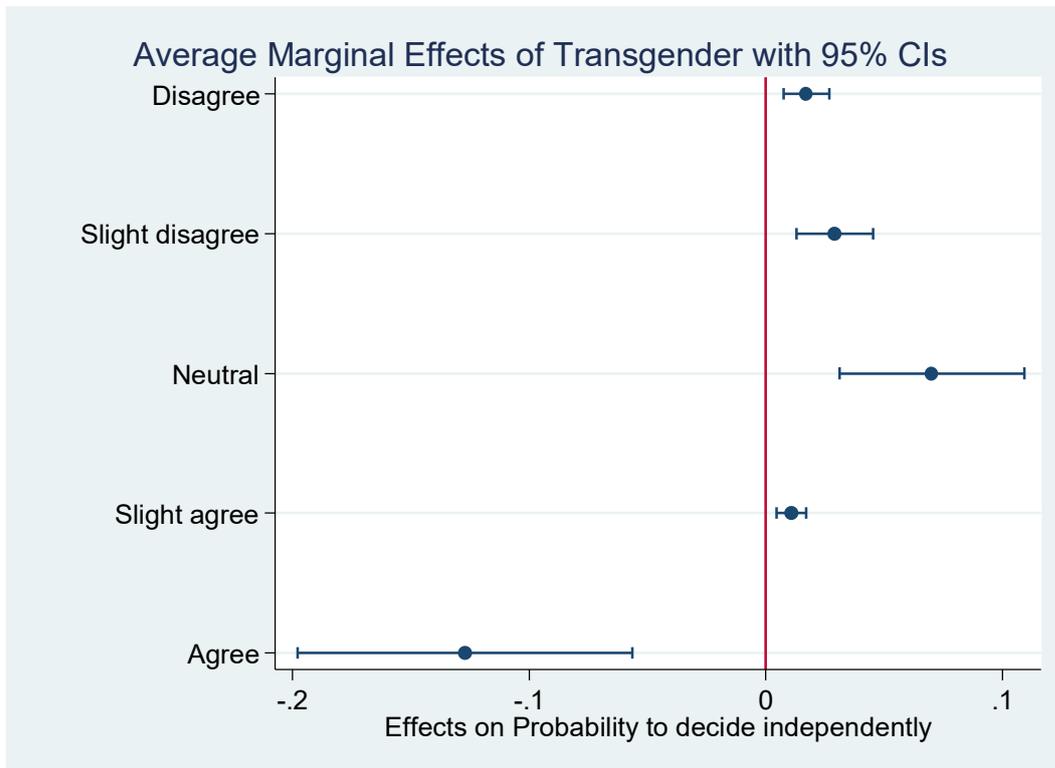

Fig 6 (a).

Response to the statement "Decisions regarding higher education, employment, etc. have been made independently and on one's own initiative."

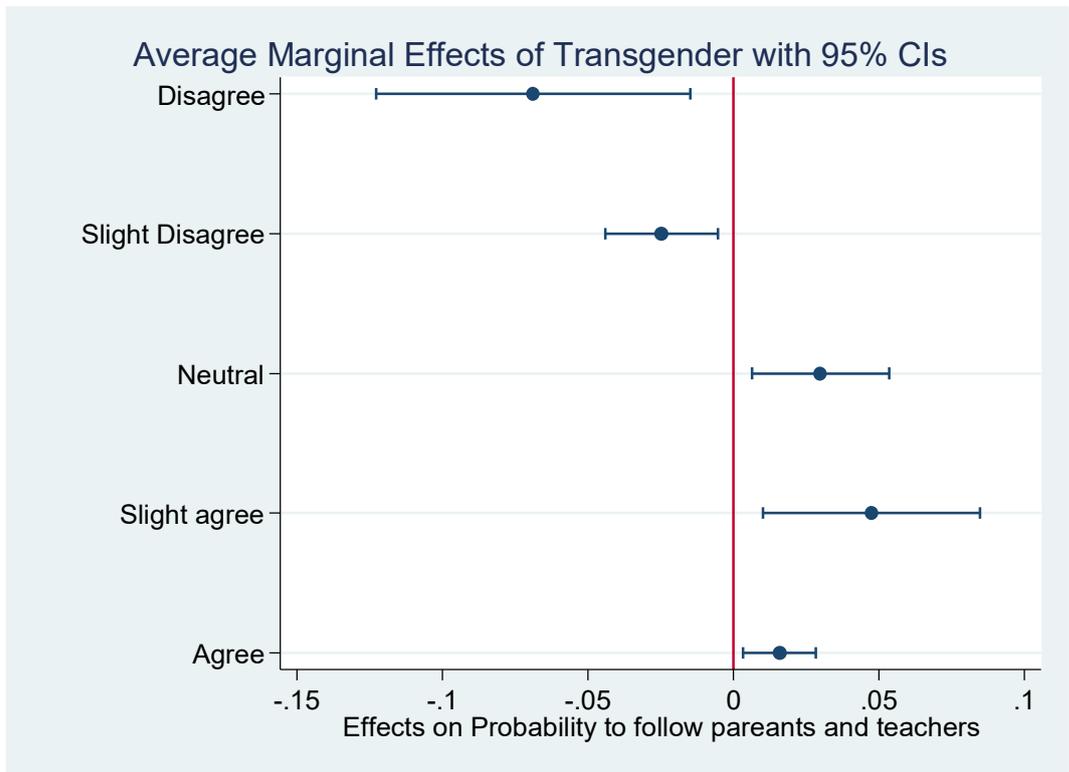

Fig 6 (b).

Response to the statement "I have followed the opinions of my parents and teachers in making decisions about higher education, employment, etc."

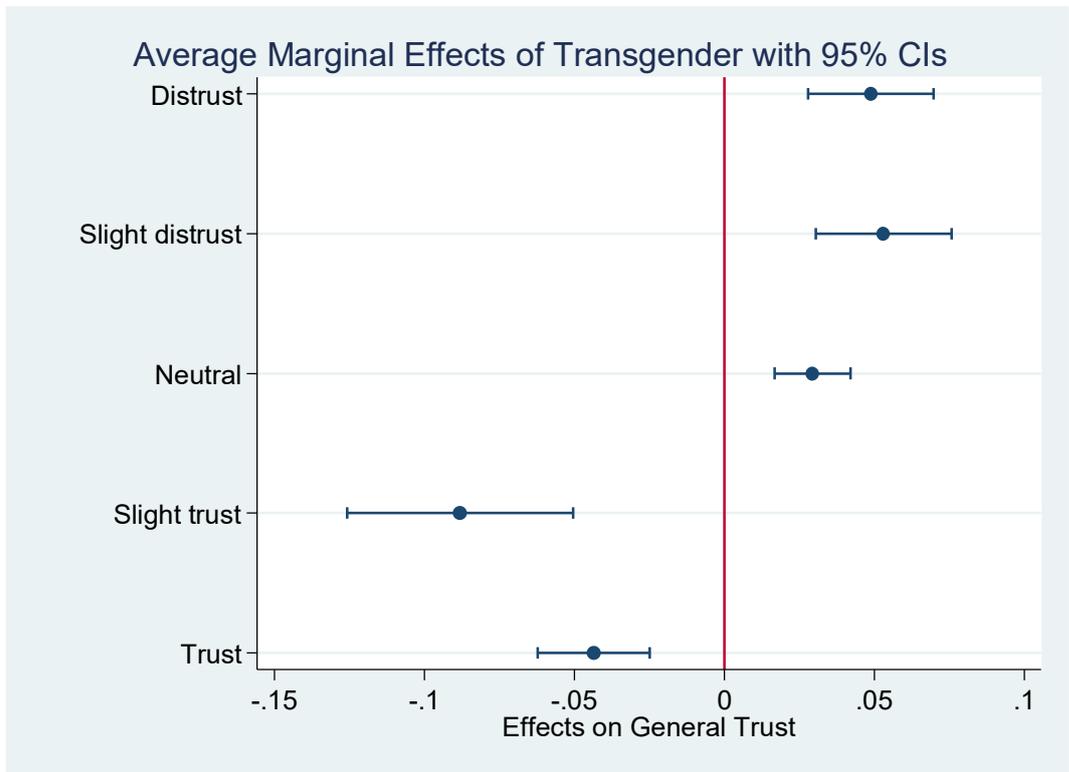

Fig7. Influence of TRANSGENDER on subjective general trust

Response to the view "People are generally trustworthy."